\documentclass[11pt,a4paper]{article}
\usepackage{amsfonts}
\usepackage{graphicx}
\usepackage{amsmath}
\usepackage{hyperref}
\usepackage{enumerate}
\usepackage{amsmath,amssymb}
\usepackage{slashed,mathrsfs}
\usepackage{feynmp,subfigure}
\usepackage{verbatim,graphicx}
\usepackage[sub,ovp]{psfragx}
\usepackage{overpic}

\RequirePackage{mathrsfs} \RequirePackage[sc]{mathpazo}
\RequirePackage{wasysym} \RequirePackage{setspace}
\newcommand{\be}{\begin{equation}}
\newcommand{\ee}{\end{equation}}
\newcommand{\bea}{\begin{eqnarray}}
\newcommand{\eea}{\end{eqnarray}}
\input{tcilatex}
\begin{document}

\date{}
\title{ \rightline{\mbox{\small
{}}}\textbf{On 750} \textbf{GeV} \textbf{Diphoton} \textbf{Resonance in
Stringy Standard-Like Models }{\ }}
\author{Adil Belhaj$^{1}$\thanks{%
belhaj@unizar.es}, Salah Eddine Ennadifi$^{2}$\thanks{%
ennadifis@gmail.com} \\
\\
{\small $^{1}$ LIRST, Facult\'{e} Polydisciplinaire, Universit\'{e} Sultan
Moulay Slimane, B\'{e}ni Mellal, Morocco } \\
{\small $^{2}$LHEP-MS, Facult\'{e} des sciences de Rabat, Universit\'{e}
Mohamed V, Rabat, Morocco} \\
}
\maketitle

\begin{abstract}
The LHC diphoton excess $750$ $GeV$ is discussed in a string-inspired
standard-like model. Precisely, a singlet scalar-extended SM from a vacua of
four stacks of intersecting D6-branes giving rise to a large gauge symmetry
is considered. Besides its relevant couplings to the SM sector, the involved
scales allow for a scalar mass near to the reported diphoton excess.


\end{abstract}

\newpage

\section{Introduction}

Recently, ATLAS and CMS collaborations have reported an excess of
diphotons associated with $750$ $GeV$ from LHC Run-II with $pp$
collisions at 13 $TeV$ \cite{1,2}. This event has received a big
interest using different methods and approaches including analytical
and simulating ones providing possible physical interpretations.
More precisely, several interpretations have been proposed based on
extensions of the standard model physics
(SM)\cite{3,4,40,5,6,7,700,701,702}. One of the most important aims
of the investigation is to propose the existence of spinless singlet
particles coupled to SM fields. Indeed, the obtained resonance could
be interpreted as a scalar field $S$ having a mass around $750$
$GeV$. The process for producing two photons are mostly generated by
two possible ways involving the fusion of either the
gluons $gg\rightarrow S\rightarrow \gamma \gamma $ or the quarks $%
qq\rightarrow S\rightarrow \gamma \gamma $. The couplings of such a
scalar could be effectively described by the terms $\mathbf{\zeta }$
$\supset $ $ \left( S/\Lambda \right) \left( G_{\mu \nu }^{a}G^{a\mu
\nu }+F_{\mu \nu }F^{\mu \nu }\right) $, where $F_{\mu \nu }$ and
$G_{\mu \nu }$ are the strong and electromagnetic fields
respectively, giving predictions on the corresponding excess and
possible production channels.

In string theory, it has been suggested that effective field theory
models can be explored to give a possible interpretation of this new
physics \cite{70,71,72,73,74,75}. This suggestion has been supported
by the fact that the important particle physics ingredients can be
embedded in intersecting D-brane models built from the orientifold
compactifications of type II superstring models. In this scenario,
the gauge groups arise from stacks of D-branes that fill 4D
spacetime and wrap appropriate cycles in the Calabi-Yau 3-folds.
However, the matter fields live at the intersection of two different
D-brane stacks in such a compactification associated with
intersection numbers which are subject to restrictions of additional
global U(1)'s\ exhibited by the orientifold compactification. In
this sens, the stringy effects, which generate corrections to the
superpotential by inducing missing couplings relevant for fermion
masses, provide an acceptable effective low-energy description for
standard-like models or some extensions \cite{8,9,10,11,12,13}. Such
models usually are pictured as intersecting lines, encoding the
gauge symmetry and matter content. These graphs allow for the
exploration of several physical effects without the need of a string
defined model. Concretely, the possible interaction coupling terms
can be derived by examining quantum numbers associated with these
brane pictures. Such a method offers a rich discussion in string
phenomenology and it can bring new feature on the corresponding
physics from higher dimensional supergravity
theories\cite{14,15,16,17,18}.

This paper aims to contribute to this issue by addressing the
observed $750$ $GeV$ diphoton excess in a stringy scalar-extended SM
based on intersecting D-brane models. Concretely, we consider a
gauge theory based on a vacua of four stacks of intersecting
D6-branes wrapping 3-cycles on the orientifold compactification with
$\mbox{U(3)}\times \mbox{Sp(1)}\times \mbox{U(1)}\times \mbox{U(1)}$
gauge symmetry. In this standard-like model, a singlet scalar $S$ is
introduced to generate, together with the standard Higgs doublet
$H$, the SM particle masses. The VEV $\left \langle S\right \rangle
$ and the mass scales $m_{s}$ of the involved new scalar offer a
possible interpretation of the diphoton excess at $750$ $GeV$
according to the known data.

The paper is organized as follows. In section 2, we build a gauge theory
based on a vacua of four stacks of intersecting D6-branes wrapping 3-cycles
on the orientifold compactification with $\mbox{U(3)}\times \mbox{Sp(1)}%
\times \mbox{U(1)}\times \mbox{U(1)}$ gauge symmetry. In section 3, we
present a possible interpretation of the diphoton excess at $750$ $GeV$ in a
singlet extension of the SM. In section 4, we approach the involved high
scales and probe the $750$ $GeV$ resonance in terms of the new scalar mass $%
m_{s}$. The last section is devoted to concluding remarks.

\section{D6-brane standard-like model}

Given that the recently LHC reported diphoton excess corresponds to a heavy
mass of $750$ $GeV$ and that likely related to a new scalar beyond the SM,
we assume that this high resonance has a stringy orgin from of a low-scale $%
M_{s}\ll M_{Planck}$ effect, and it is a singlet scalar under the SM gauge
group. We thus build a stringy model based on four stacks of D6-branes with
a flavor symmetry distinguishing various quarks from each others. In this
brane building, the gauge symmetry is
\begin{equation}
\mbox{U}(3)_{a}\times \mbox{Sp}(1)_{b}\times \mbox{U}(1)_{c}\times \mbox{U}%
(1)_{d},  \label{eq1}
\end{equation}%
where the $\mbox{Sp}(1)\simeq \mbox{SU}(2)$ weak symmetry arises from
D6-wrapped on an orientifold invariant three-cycle ($b=b^{\ast }$). It has
been shown that there is no difference between the quark doublets since they
have all the same $\mbox{U}(1)_{a,c,d}$ charges. A close inspection shows
that one should consider a D6-brane configuration obtained from the
compactification of type IIA superstring on $\Pi _{i=1}^{3}T_{i}^{2}$. The
intersection numbers can be obtained from the wrapping numbers of the
D6-branes around the $T^{2}$ factors. An appropriate choice of such numbers
gives the following intersections

\begin{table}[th]
\label{t:one}
\par
\begin{center}
\begin{tabular}{|c|c|c|c|c|c|c|c|}
\hline
$Sector$ & $ab$ & $ac$ & $ac^{\ast }$ & $ad$ & $ad^{\ast }$ & $db$ & $%
dc^{\ast }$ \\ \hline
$Inter\sec tion$ & $3$ & $-2$ & $-1$ & $-1$ & $-2$ & $3$ & $-3$ \\ \hline
\end{tabular}
\label{t:one}
\end{center}
\caption{An intersection numbers of the SM spectrum. The other ones are set
to zero.}
\end{table}
The field content and the corresponding charges are illustrated in the table
1.
\begin{table}[th]
\label{t:one}
\par
\begin{center}
\begin{tabular}{|c|c|c|c|c|c|c|c|c|}
\hline
$Sector$ & $ab$ & $ac$ & $ac^{\ast }$ & $ad$ & $ad^{\ast }$ & $db$ & $%
dc^{\ast }$ & $bc$ \\ \hline
$Fields$ & $q^{i}$ & $\overline{u}^{2,3}$ & $\overline{d}^{3}$ & $\overline{u%
}^{1}$ & $\overline{d}^{1,2}$ & $\ell ^{i}$ & $\overline{e}^{i}$ & $H$ \\
\hline
$Rep$ & $3(3,\overline{2})$ & $2(\overline{3},1)$ & $1(\overline{3},1)$ & $1(%
\overline{3},1)$ & $2(\overline{3},1)$ & $3(1,\overline{2})$ & $3(1,1)$ & $%
1(1,2)$ \\ \hline
$Q_{a}$ & $1$ & $-1$ & $-1$ & $-1$ & $-1$ & $0$ & $\ 0$ & $0$ \\ \hline
$Q_{c}$ & $0$ & $1$ & $-1$ & $0$ & $0$ & $0$ & $-1$ & $1$ \\ \hline
$Q_{d}$ & $0$ & $0$ & $0$ & $1$ & $-1$ & $1$ & $-1$ & $0$ \\ \hline
$Y$ & $1/6$ & $-2/3$ & $1/3$ & $-2/3$ & $1/3$ & $-1/2$ & $1$ & $-1/2$ \\
\hline
\end{tabular}
\label{t:one}
\end{center}
\caption{SM spectrum and their U(1)$_{a,c,d}$ charges for $Y=\frac{1}{6}%
Q_{a}-\frac{1}{2}Q_{c}-\frac{1}{2}Q_{_{d}}$. The index $i=1,2,3$ is the
family index.}
\end{table}
In this construction, the three left-handed quarks $q^{i}$ are localized at
intersections of D6-branes $a$ and $b$ while right-handed quarks, $\overline{%
u}^{i}$ and $\overline{d}^{i}$ split into two up quarks $\overline{u}^{2,3}$
and one down quark $\overline{d}^{3}$ which are localized at intersection of
the D6-branes $a$ and $c/c^{\ast }$. Two down quarks $\overline{d}^{1,2}$
and one up quark $\overline{u}^{1}$ are placed at intersection of the
D6-branes $a$ and $d/d^{\ast }.$ The three left-handed leptons $\ell ^{i}$
arise at the intersection of branes $b$ and $d/c,c^{\ast }$ respectively.
Moreover, the three right-handed electrons $\overline{e}^{i}$ arise at the
intersection of D6-branes $d$ and $c^{\ast }.$ Finally, the Higgs doublet $H$
arises at the intersection of D6-branes $b$ and $c/c^{\ast }$. The origin of
the matter fields is associated with the linear combination of $\mbox{U}%
(1)_{a,c,d}$ reproducing the SM particle hypercharges.

The corresponding model can be illustrated in figure 1.

\begin{center}
\begin{figure}[th]
\begin{center}
{\includegraphics[width=10cm]{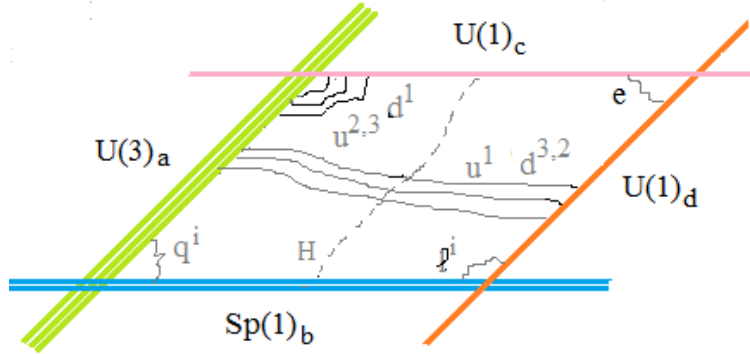}}
\end{center}
\caption{Four-Stack Stringy SM. Bold lines denote D6-branes and thin lines
denote chiral and scalar spectrum.}
\end{figure}
\end{center}

The 4D Yukawa coupling terms can be derived with respect to the symmetry
charges proposed in table 1. In fact, the $\mbox{U(1)}_{a,c,d}$ field
charges can be used to write down the possible Yukawa couplings interaction
terms for the heavy quarks and leptons. These terms are

\begin{eqnarray}
\mathbf{Q}_{c,d}\left( H^{\dagger }q\overline{c}\right) &=&Q_{c,d}\left(
H\right) +Q_{c,d}\left( q\right) =Q_{c,d}\left( \overline{c}\right) =0,
\notag \\
\mathbf{Q}_{c,d}\left( H^{\dagger }q\overline{t}\right) &=&0,  \notag \\
\mathbf{Q}_{c,d}\left( Hq\overline{b}\right) &=&0,  \label{eq2} \\
\mathbf{Q}_{c,d}\left( H\ell ^{i}\overline{e}^{i}\right) &=&0.  \notag
\end{eqnarray}%
The corresponding Lagrangian is then
\begin{equation}
\mathbf{\zeta }_{Yuk}=y_{c}H^{\dagger }q\overline{c}+y_{t}H^{\dagger }q%
\overline{t}+y_{b}Hq\overline{b}+y_{e^{i}}H\ell ^{i}\overline{e}^{i},
\label{eq3}
\end{equation}%
where $y^{\prime }s\leqslant 1$ are coupling constants accounting for the
Higgs-fermion interaction strenghts between these terms.

\section{ Singlet scalar-extension}

It turns out that the remaining phenomenologically desired coupling terms
can be implemented by U(1)'s charged scalars extendeding the SM \cite%
{14,15,16}. In type IIA superstring theory, a scalar can be obtained from
either the geometric deformation or the stringy deformation associated with
NS-NS and R-R fieldd on cycles of internal manifolds. In what follows, we
add a siglet complex scalar field $S$ obtained by combining a complex
structure deformation and the R-R 3-form wrapping a 3-cycle in the internal
space. The reason for adding such a scalar is to generate missing coupling
terms with respect to U(1) charges. The new scalar must have the charges,

\begin{table}[th]
\begin{center}
\begin{tabular}{|c|c|c|c|c|c|c|}
\hline
$Sector$ & $Field$ & $Rep$ & $Q_{a}$ & $Q_{c}$ & $Q_{d}$ & $Y$ \\ \hline
$cd$ & $S$ & $1(1,1)$ & $\ 0$ & $1$ & $-1$ & $0$ \\ \hline
\end{tabular}%
\end{center}
\caption{Required extra singlet with its U(1)$_{c,d}$ charges.}
\end{table}

This is shown in the figure 2.

\begin{center}
\begin{figure}[th]
\begin{center}
{\includegraphics[width=8cm]{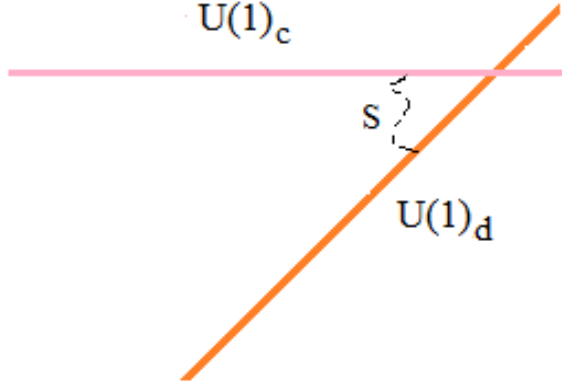}}
\end{center}
\caption{New scalar and its associated U(1)$_{c,d}$ charges indicated by the
dotted thin line.}
\end{figure}
\end{center}

With this scalar adding, the absent terms are now generated via higher order
terms and thus will be suppressed\ by factors $\left \langle S\right \rangle
^{n}/M_{s}^{m}$, where $n$ and $m$ are power integer numbers, $M_{s}$
denotes the string mass and $\left \langle S\right \rangle $ its VEV. These
terms are now

\begin{eqnarray}
\mathbf{Q}_{c,d}\left( SH^{\dagger }q\overline{u}\right)  &=&Q_{c,d}\left(
S\right) +Q_{c,d}\left( H\right) +Q_{c,d}\left( q\right) =Q_{c,d}\left(
\overline{u}\right) =0,  \notag \\
\mathbf{Q}_{c,d}\left( S^{\ast }Hq\overline{d}\right)  &=&0,  \notag \\
\mathbf{Q}_{c,d}\left( S^{\ast }Hq\overline{s}\right)  &=&0,  \label{eq4} \\
\mathbf{Q}_{c,d}\left( S^{2}(H\ell ^{i})^{2}\right)  &=&0.  \notag
\end{eqnarray}%
The corresponding Lagrangian is then
\begin{eqnarray}
\mathbf{\zeta }_{Yuk}^{\prime } &=&M_{s}^{-1}\left( y_{u}SH^{\dagger }q%
\overline{u}+y_{d}S^{\ast }Hq\overline{d}+y_{s}S^{\ast }Hq\overline{s}%
\right)   \notag \\
&&+M_{s}^{-3}y_{\nu ^{i}}S^{2}(H\ell ^{i})^{2}.  \label{eq5}
\end{eqnarray}%
In this model, the VEV $\left \langle S\right \rangle $ induces the
perturbatively missing Yukawa couplings and along with the Higgs VEV the
masses for these light fermions (\ref{eq5}). Compared to the previous
contributions given in (\ref{eq3}), they are suppressed by the string mass
scale $M_{s}{}$ with a high suppression for the left-handed neutrino terms.

\section{Probe of the 750 GeV mass}

Without requiring geometric specifics of the defined stringy model, more
results beyond the interaction terms could be derived. Particulary, it
includes the involved string scale $M_{s}$ with the VEV $\left \langle
S\right \rangle $ and mass $m_{s}$ of the scalar field $S$. For that, after
the electroweak symmetry breaking by the Higgs VEV at $\left \langle
H\right
\rangle \simeq 246$ $GeV$, adequate combinations of fermion masses,
for which their net scalar-fermion couplings could be absorbed, can give
approximate values of the new scales. Indeed, using the left-handed neutrino
mass terms appearing in (\ref{eq5}) with an upper bound of $m_{\nu _{\tau
}}\lesssim 1$ $eV$, \ we can \ predict the string scale $M_{s}$ as
\begin{equation}
M_{s}=\frac{y_{\nu _{\tau }}}{y_{u}^{2}}\frac{m_{u}^{2}}{m_{\nu }}\sim
10^{4}GeV,  \label{eq6}
\end{equation}%
and then the scalar VEV $\left \langle S\right \rangle $ becomes
\begin{equation}
\left \langle S\right \rangle =\frac{y_{c}}{y_{s}}M_{s}\frac{m_{s}}{m_{c}}%
\sim 10^{3}GeV.  \label{eq7}
\end{equation}%
At this point, it is worth mentioning that we have two new high scales: one
belongs to the low-string scale $M_{s}$ (\ref{eq6}), and the other belongs
to the VEV of the new scalar $\left \langle S\right \rangle $ given in eq.( %
\ref{eq7}). Other than the partial explanation of the fermion mass
hierarchies and the smallness of neutrino masses, these net scales allow for
the possibility to wonder if the recent $750$ $GeV$ resonance is somehow
related to such low-scale stringy effect. In fact, we can go further and
calculate the mass of the new scalar $S$. The latter is proportional to the
scalar VEV given in eq.(\ref{eq7}) through its self-coupling parameter $%
\lambda _{s-s}\leqslant 1$ such as
\begin{equation}
m_{S}=\sqrt{\lambda _{s-s}}\left \langle S\right \rangle \leqslant 10^{3}GeV,
\label{eq8}
\end{equation}%
where we see clearly that for the scalar self-coupling value $\lambda
_{s-s}\simeq 0,562$, one can get a mass of $m_{S}\simeq m_{\gamma \gamma
}=750GeV$ corresponding to the reported LHC diphoton excess. These new
scales that result is a stringy prediction for physics beyond SM push to ask
whether the present LHC Run II at $\sqrt{s}=13$ $TeV$ is able to see more
significant stringy physics directly.

\section{Conclusion and related remarks}

In this work, we have discussed the LHC diphoton excess at $750$ $GeV$ in a
string-inspired gauge theory where the corresponding effective low-energy
theory emerges from type IIA superstring on the orientifold
compactification. Concretely, we have considered four stacks of intersecting
D6-branes configuration producing the SM spectrum extended by a singlet
scalar $S$. The corresponding effective superpotential produces different
coupling scales relative to allowed perturbative and higher order suppressed
terms. Attributing the allowed perturbative terms to the known heavy quarks
and leptons and the higher order generated terms to known light quarks and
neutrinos, the hierarchy of fermion masses find an a explanation through
higher order terms suppressed\ by factors $\left
\langle S\right \rangle
^{n}/M_{s}^{m}$. We have further probed the stringy explanation of LHC $750$
$GeV$ diphoton excess by probing the new scales involved in the model,
low-string scale $M_{s}$ $\sim 10^{4}GeV$ and the scalar VEV $\left \langle
S\right \rangle \sim 10^{3}GeV$, by referring to known data and then we have
shown how the new scalar can have a mass of the detected diphoton excess $%
m_{S}$ $\simeq m_{\gamma \gamma }=750$ $GeV$.\newline
Alternatively gauge theories can be geometrically engineered from
singularities of the K3 surface \cite{19}. A way to get such gauge theories
from type IIA superstring is to consider a compactification on a singular
3-fold with K3 surface fibration over a base space B. In this way, the gauge
symmetry and matter fields are obtained from the singularities of the fiber
and the non-trivial geometry of the base space, respectively. In connection
with the model presented here, it is possible to consider an SU(7)
singularity in the K3 fiber. Thus, the deformation of such a singularity
could produce the following decomposition $\mbox{SU(7)}\rightarrow %
\mbox{SU(3)}\times \mbox{SU(2)}\times \mbox{U(1)}\times \mbox{U(1)}$ gauge
symmetry. It would also be interesting to explore the geometric engineering
method based on string compactifications on singular manifolds. We hope to
report elsewhere on this possible connection.\newline
\textbf{Acknowledgements}: The authors would like to deeply thank Maria
Pilar Garcia del Moral for discussions on related topics.


\begin{thebibliography}{99}
\bibitem{1} The ATLAS collaboration, ATLAS-CONF-2015-081.

\bibitem{2} CMS Collaboration [CMS Collaboration], CMS-PAS-EXO-1 5-004.

\bibitem{3} A. Falkowski, J. F. Kamenik, \emph{Di-photon portal to warped
gravity}, \texttt{arXiv:1603.06980}.

\bibitem{4} M. Bauer, C. Hoerner, M. Neubert, \texttt{arXiv:1603.05978}.

\bibitem{40} R. Benbrik, C. H. Chen, T. Nomura, Phys. Rev. \textbf{D93}
(2016)055034 .

\bibitem{5} T. Li, J. A. Maxin, V. E. Mayes, D. V. Nanopoulos, \texttt{%
arXiv:1602.00949}.

\bibitem{6} S. F. King, R. Nevzorov, \texttt{arXiv:1601.07242}.

\bibitem{7} C. Hati, \texttt{arXiv:1601.02457}.

\bibitem{700} F. Staub, P. Athron, L. Basso, M. D. Goodsell,
D. Harries, M. E. Krauss, K. Nickel, T. Opferkuch, L.  Ubaldi,
 A. Vicente, A. Voigt, {\em Precision tools and models to narrow in on the 750 GeV diphoton
 resonance}, {\tt arXiv:1602.05581 }.
 \bibitem{701} P. Roig, J.J. Sanz-Cillero, {\em
A broad 750 GeV diphoton resonance? Not alone},  {\tt
arXiv:1605.03831}.
\bibitem{702} P. S. B. Dev, R. N. Mohapatra, Y. Zhang
Q. Seesaw, {\em  Vectorlike Fermions and Diphoton Excess}, JHEP{\bf
02},  (2016)186,  { \tt arXiv:1512.08507}.
\bibitem{70} J. Heckman, \emph{750 GeV Diphotons from a D3-brane}, \texttt{%
arXiv:1512.06773}.

\bibitem{71} M. Cvetic, J. Halverson, P. Langacker, \emph{String
Consistency, Heavy Exotics, and the $750$ GeV Diphoton Excess at the LHC},
\texttt{arXiv:1512.07622}.

\bibitem{72} L. E. Ibanez, V. Martin-Lozano, \emph{A Megaxion at 750 GeV as
a First Hint of Low Scale String Theory}, \texttt{arXiv:1512.08777}.

\bibitem{73}
  T.~Li, J.~A.~Maxin, V.~E.~Mayes and D.~V.~Nanopoulos, {\em The $750$ GeV Diphoton Excesses in a Realistic D-brane
  Model}, {\tt
  arXiv:1602.09099 [hep-ph]}.
  \bibitem{74}
L. A. Anchordoqui, I. Antoniadis, H. Goldberg, X. Huang, D. Lust, T.
R.  Taylor, {\em Update on 750 GeV diphotons from closed string
    states},  Phys.Lett. {\bf B759} (2016) 223-228 {\tt    arXiv:1603.08294}.
\bibitem{75}
    L.  A. Anchordoqui, I.  Antoniadis, H. Goldberg, X. Huang, D. Lust, T.  R.
    Taylor, {\em 750 GeV diphotons from closed string states},  Phys.Lett. {\bf B755} (2016) 312-315, {\tt
    arXiv:1512.08502}.


\bibitem{8} G. Aldazabal, L. E. Ib\'{a}\~{n}ez, F. Quevedo, J. High Energy
Phys. \textbf{02}(2000) 015.

\bibitem{9} P. Anastasopoulos, T. P. T. Dijkstra, E. Kiritsis, A. N.
Schellekens, Nucl. Phys. \textbf{B759} (2006)83.

\bibitem{10} R. Blumenhagen, M. Cvetic, P. Langacker, G. Shiu, Ann. Rev.
Nucl. Part. Sci. \textbf{71} (2005) 55.

\bibitem{11} L. E. Ib\'{a}\~{n}ez, F. Marchesano, R. Rabadan, J. High Energy
Phys. \textbf{11} (2001) 002.

\bibitem{12} M. Cvetic, G. Shiu, A. Uranga, Nucl. Phys. \textbf{B615}
(2001)3.

\bibitem{13} M. Cvetic, I. Papadimitriou, Phys. Rev. \textbf{D67}
(2003)126006.

\bibitem{14} D.V. Gioutsos, G.K. Leontaris, J. Rizos, Eur. Phys. J. \textbf{%
C45} (2006) 241.

\bibitem{15} D. Cremades, L. E. Ib\'{a}\~{n}ez, F. Marchesano, J. High
Energy Phys. \textbf{07} (2003) 038.

\bibitem{16} M. Cvetic, J. Halverson, R. Richter, \texttt{arXiv:0909.4292}.

\bibitem{17} C. Panagiotakopoulos, K. Tamvakis, Phys. Lett. \textbf{B145}
(1999) 469.

\bibitem{18} A. Belhaj, M. Benhamza, S.E. Ennadifi, S. Nassiri, E.H. Saidi,
\emph{On Fermion Mass Hirerachies in MSSM-like Quiver Models with Stringy
Correct ions}, Cent. Eur. J. Phys, \textbf{9} (2011)1458, \texttt{%
arXiv:1107.0872}.

\bibitem{19} Katz, C. Vafa, \emph{Matter from geometry}, Nucl. Phys. \textbf{%
B497} (1997) 146, \texttt{hep-th/9606086}.
\end{thebibliography}
\end{document}